\documentclass{aa}
\usepackage[dvips]{graphicx}
\usepackage{natbib}
\bibpunct{(}{)}{;}{a}{}{,}
\usepackage{txfonts}
\usepackage[mathscr]{eucal}
\begin{document}
\title{Near-infrared polarimetric study of the bipolar nebula\\
IRAS 19312+1950
}

\author{
	K. Murakawa\inst{1},
        J. Nakashima\inst{2},
	K. Ohnaka\inst{1},
	\and
	S. Deguchi\inst{3}
	}

\offprints{Koji Murakawa, \email{murakawa@mpifr-bonn.mpg.de}}

\institute{
	Max-Planck-Institut f\"{u}r Radioastronomie,
        Auf dem H\"{u}gel 69, D-53121 Bonn, Germany
        \and
        Academia Sinica Institute of Astronomy and Astrophysics,
        P.O. Box 23-141, Taipei, Taiwan
	\and
        Nobeyama Radio Astronomy, National Astronomical Observatory
        of Japan, Minamimaki, Minamisaku, Nagano 384-1305, Japan
	}


\abstract{}
%
{
We have investigated the properties of the central star and dust in the bipolar
nebula IRAS~19312+1950, which is an unusual object showing characteristics of
a supergiant, a young stellar object, and an asymptotic giant branch (AGB) star.
}
%
{
We obtained $H$-band polarimetric data of IRAS~19312+1950 using the
near-infrared camera (CIAO) on the 8~m Subaru telescope. In order to
investigate the physical properties of the central star and the nebula,
we performed dust radiative transfer modeling and compared the model results
with the observed spectral energy distributions (SEDs), the radial profiles
of the total intensity image, and the fraction of linear polarization map.
}
%
{
The total intensity image shows a nearly spherical core with $\sim$3$\arcsec$
radius, an \textsf{S}-shaped arm extending $\sim$10$\arcsec$ in the northwest
to southeast direction, and an extended lobe towards the southwest.
The polarization map shows a centro-symmetric vector alignment in almost the
entire nebula and low polarizations along the \textsf{S}-shaped arm. These
results suggest that the nebula is accompanied by a central star, and the
\textsf{S}-shaped arm has a physically ring-like structure. From our radiative
transfer modeling, we estimated the stellar temperature, the bolometric
luminosity, and the current mass-loss rate to be 2\,800~K, $7\,000~L_{\sun}$,
and $5.3\times10^{-6}~M_{\sun}$yr$^{-1}$, respectively.
}
%
{
Taking into account previous observational results, such as the detection
of SiO maser emissions and silicate absorption feature in the 10~$\mu$m
spectrum, our dust radiative transfer analysis based on our near-infrared
imaging polarimetry suggests that (1) the central star of IRAS~19312+1950 is
likely to be an oxygen-rich, dust-enshrouded AGB star and (2) most of the
circumstellar material originates from other sources (e.g. ambient dark clouds)
rather than as a result of mass loss from the central star.
}

\keywords{near-infrared -- imaging polarimetry -- radiation transfer modeling
-- individual (\object{IRAS~19312+1950}) -- (post-)AGB stars -- circumstellar matter}

\titlerunning{H-band imaging polarimetry of IRAS~19312+1950}
\authorrunning{Murakawa et al.}
\maketitle

\section{Introduction}

\object{IRAS~19312+1950} is a peculiar bipolar nebula with a horn-like
appearance extending $\sim$$30\arcsec$ in the 2MASS $JHK$-band images.
With IRAS colors $[12]-[25]=0.50$ and $[25]-[60]=0.71$, it falls into region
VIII on the IRAS color-color diagram \citep{veen88}. A relatively large
fraction of objects in this region is known to belong to different types such
as Miras, OH$/$IR stars, or carbon stars. A weak 10~$\mu$m absorption was
detected in the IRAS Low-Spectral Resolution spectra of IRAS~19312+1950
\citep{volk91}. Maser emissions of silicon monoxide (SiO), water (H$_2$O)
\citep{nd00}, and hydroxyl (OH) (Lewis,~B.~M., private communication) were also
detected. These observational results led the authors to conclude that
IRAS~19312+1950 is a candidate for an oxygen-rich asymptotic giant branch (AGB)
or post-AGB star \citep{nd00}. However, some unusual characteristics have been
found in subsequent observations, and debates on the true nature of
IRAS~19312+1950 have arisen. \citet{deguchi04} found C-, N- and O-bearing
molecules (e.g.~H$^{13}$CN, CH$_3$OH, SO and SO$_2$) towards this object.
C- and N-bearing molecules such as H$^{13}$CN and CH$_3$OH are often detected
in carbon-rich circumstellar environments or in molecular cloud cores in a dark
cloud. While C-bearing molecules and SO take $\sim$$10^5$~yrs and
$\sim$$10^6$~yrs, respectively, to be formed in an environment like a dark
cloud, the time scale ($\sim$$10^4$~yrs) of expanding gas with an expansion
velocity of 10~km~s$^{-1}$ to cross the envelope of IRAS~19312+1950
\citep[$\sim$5.8$\times10^{17}$~cm at 2.5~kpc,][]{nakashima04} is remarkably
shorter than the above time scales. Furthermore, a stellar velocity of
$\sim$36~km~s$^{-1}$ was found from the velocity components of the SiO and
H$_2$O masers. Besides the broad component in the thermal emission of some
molecules, the presence of a narrow component in the thermal emission of CO,
HCN, CS, SO and HCO$^+$ suggests the presence of an additional, kinematically
cold component, which has been interpreted to originate from a dust cloud rather
than the matter ejected by mass loss from the central star
\citep{nd00,deguchi04}. In addition, \citet{deguchi04} estimated the envelope
mass and the mass-loss rate to be
$\sim$25~$M_{\sun}$ and $\sim$$2.6\times10^{-4}~M_{\sun}$yr$^{-1}$,
respectively, by means of the large-velocity gradient (LVG) model with their
observational results of several molecular lines in the radio frequency domain.
For the above reasons, to date, it seems plausible that IRAS~19312+1950 is a
progenitor with an initial mass of $M_\star\ge4~M_{\sun}$ and is embedded
in an ambient cloud by chance \cite[see also][]{nd05}.

Previous studies of this object have mainly focused on identifications of
gas-phase molecules and investigations of their kinematics. In order to provide
a better understanding of this object's nature, model analyses to explain
observed spectral energy distributions (SEDs) and images as well as to
investigate the stellar parameters are important. Although dust is expected to
contribute only $\sim$1~\% of the mass in the envelope \citep[e.g][]{knapp74},
it plays a more important role in the transfer of radiation from the optical to
far-infrared (FIR) than gas. Furthermore, light scattered by dust in the
envelope often dominates in the SED at optical and near-infrared (NIR)
wavelengths, and the maximum dust grain size in envelopes around AGB stars and
post-AGB stars is expected to be on the order of 0.1~$\mu$m
\citep[e.g.][]{jura96}. Polarization analysis in the NIR is a powerful
technique for probing such a dusty environment. In our experiments, therefore,
we performed imaging polarimetry of the IRAS~19312+1950 nebula in the NIR and
radiative transfer modeling to derive the stellar parameters and dust
properties in the envelope by comparing these with the observed SEDs, the total
intensity image, and the polarization properties. The results of our
observations and numerical simulations are presented in sect.\,\ref{observation}
and \ref{modeling}, respectively. In sect.\,\ref{discussion}, we discuss the
properties of dust in the envelope and the central star of IRAS~19312+1950.

\section{Observations and Results}\label{observation}
\subsection{$H$-band imaging polarimetry}
We obtained polarimetric images of IRAS~19312+1950 using the NIR camera (CIAO)
on the 8~m Subaru telescope on 2 August 2002. Because the object is faint in
the $J$-band ($m_J=11.3$~mag) and the nebulosities are easier to detect in the
$H$ band than the $K$-band, we only tried the $H$-band in this experiment.
The central wavelength and the band width of the $H$-band filter are
$\lambda_\mathrm{c}=1.65~\mu$m and $\Delta\lambda=0.30~\mu$m, respectively.
We used the medium resolution camera with a 0.0217~arcsec~pix$^{-1}$ pixel
scale ($22\arcsec\times22\arcsec$ field of view (FOV)). Since the target is too
faint at the wavelength for adaptive optics (AO) wavefront sensing
($m_R$=18~mag), the AO was not used. The natural seeing was $\sim$0\farcs6 at
$H$-band. We followed the observing sequence described in our previous paper
\citep{murakawa05}. In order to measure linear polarization, we obtained four
sets of images with orientations of the half-wave plate (HWP) of $0\degr$,
$45\degr$, $22.5\degr$, and $67.5\degr$. The exposure time for each frame
acquisition was 3~sec. Five frames were obtained for each orientation of the
HWP and eight dither offsets of $10\arcsec$ separation were performed.
The total integration time was 480~sec. The nebula of the target covers the
entire FOV of the camera, and complex interstellar diffuse emission and
polarization are expected \citep{axson76} because the target is located near
the galactic plane (galactic coordinate of $55\degr.37$, $+0\degr.185$).
Thus, we also obtained sky frames for calibration of the sky background level
and interstellar polarization. The position offsets are 20$\arcsec$ east and
west of the target. The same observing procedures as those of the target were
applied.

We reduced the observed data in a similar manner as described in our previous
papers \citep{murakawa04,murakawa05}. We first subtracted the dark frame and
applied flat fielding. We calibrated the sky background level and the
interstellar polarization by subtracting the sky frames. Then we obtained
the Stokes $IQU$ images, converting with formula
$I=\left(I_{0.0}+I_{22.5}+I_{45.0}+I_{67.5}\right)/2$,
$Q=\left(I_{0.0}-I_{45.0}\right)/\eta$,
and $U=\left(I_{22.5}-I_{67.5}\right)/\eta$. Where $\eta$ is the total
polarization efficiency of the HWP and the wire-grid polarizer, the value of
0.875 \citep{murakawa04} is used in this experiment. The polarized intensity
($PI$), the fraction of linear polarization ($P$), and the polarization position
angle ($\theta$) are also derived by $PI=\sqrt{Q^2+U^2}$, $P=PI/I$, and
$\theta=1/2\arctan\left(U/Q\right)$, respectively. We also obtained error images
of all Stokes parameters, the fraction of linear polarization, and the
polarization position angle. The signal-to-noise ratios of the Stokes $I$ image
are 7--10 in a region within 1$\arcsec$ from the central star and 30--60
in a region $\sim$$3\arcsec$ from the central star. The estimated background
surface brightness and the polarization in the neighboring sky are
$7\times10^{-15}$~Wm$^{-2}\mu$m$^{-1}$arcsec$^{-2}$
\citep[13.0 mag arcsec$^{-2}$, cf. 13.4 mag arcsec$^{-2}$ for the averaged
sky background level,][]{tokunaga00} and 1--2~\%, respectively. The error of
linear polarization is 2--5~\% per pixel.

\begin{figure*}
  \centering
  \includegraphics[width=17.0cm,keepaspectratio]{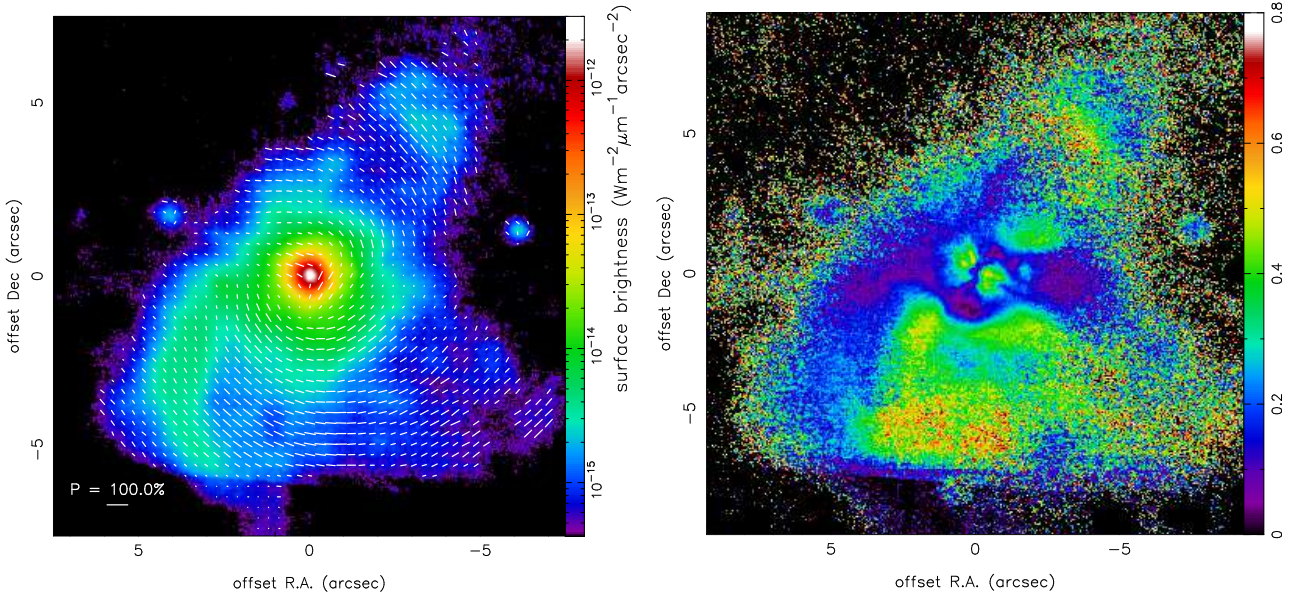}
  \caption{Results of $H$-band imaging polarimetry of IRAS~19312+1950. The field
           of view is $15\arcsec\times15\arcsec$. a) The total intensity
           (Stokes $I$) image. The polarization vectors are also plotted every
           14 pixels ($\sim$$0\farcs3$). b) The fraction of linear polarization
           map. The poor image quality $2\arcsec$ from the bottom is due to the
           lack of data in some dither positions.
          }
  \label{ciao}
\end{figure*}

\subsection{Polarization map} \label{polmap}
\begin{figure}
  \centering
  \includegraphics[width=8.0cm,keepaspectratio]{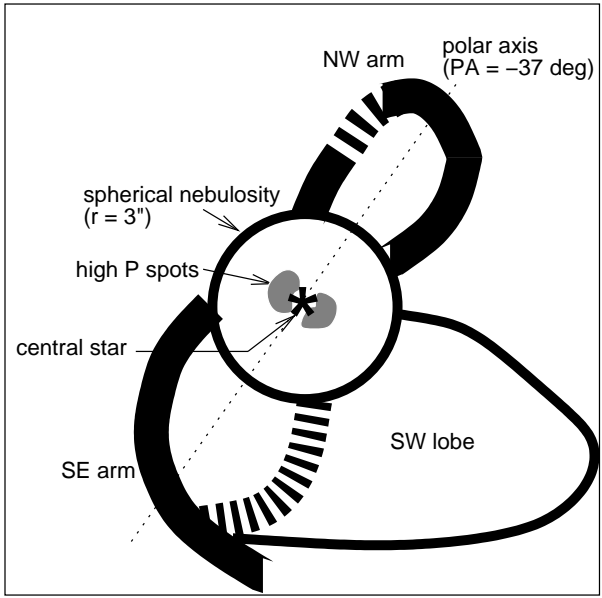}
  \caption{This cartoon explains what is detected in our images.
          }
  \label{cartoon}
\end{figure}

Figure \ref{ciao} (a) and (b) present a total intensity map (Stokes $I$) with
polarization vectors and a fraction of linear polarization map, respectively.
The field of view (FOV) is 15\arcsec$\times$15\arcsec. Figure \ref{cartoon}
shows what is seen in our images. The total intensity map shows a nearly
spherical nebulosity within $3\arcsec$ of the intensity peak, two ring-like
structures extending $\sim$10$\arcsec$ to the northwest and the southeast (PA of
$-37\degr$ with respect to the intensity peak), and a lobe extending southwest.
The ring-like structures look rather point-symmetric. These nebulous features
are seen as a horn-like structure in the 2MASS $JHK$-band images \citep{nd00}
and a NIR image \citep{deguchi04}. The polarization vectors are
centro-symmetrically aligned in the entire nebula, where a sufficient signal is
detected, and the fraction of linear polarization attains 30--60~\% in
nebulosities at NW, SE and SW. These features suggest that the nebulosity is
seen with singly scattered light in the $H$-band, which is emitted from the
central star, and the nebulae are associated with the central star.
In the fraction of linear polarization map, we see low polarization ($P<20~\%$)
in the SE and NW arms. A possible interpretation is scattering angle dependence
of polarization properties. The fraction of linear polarization has the maximum
value near the normal scattering angle, and low values in forwardly or
backwardly scattered light \citep[see Fig.~2 in the paper by][]{fischer94}.
If we assume that the \textsf{S}-shaped arm has a physically ring-like dust
structure and the plane of the ring is tilted with respect to the plane of the
sky, the scattered light from the ring is expected to have low polarizations.
Such a polarization distribution is remarkably different from IRAS~17441--2441
\citep{oppenheimer05}, in which polarization is enhanced at the rim of the
bipolar lobes. In this object, the bipolar lobes are hollow and light scattered
at the rim in the plane of the sky results in a large fraction of polarization.
The different characteristics in the polarization map of IRAS~19312+1950 are
due to a different density structure of its nebula. We also identified two
spots with high polarization within $1\arcsec$ of the central star. The spots
are aligned in the equatorial direction; i.e., perpendicular to the polar
direction. A similar signature is seen in an optically thin nebula with an
equatorially enhanced density distribution
\citep[e.g.][]{murakawa07}. It may be possible
that dust condenses more in the equatorial direction than in the polar axis.
However, it is slightly suspicious because of the natural seeing-limited
angular resolution ($\sim$$0\farcs6$ at $H$-band) in our image.

We compare our NIR image with the results of previous CO observations
\citep{nd05}. Narrow components of 35--36 and 37--38\,km\,s$^{-1}$
$\element[][13]{CO}$ $J=1$--0 emission lines extending $\sim20\arcsec$ roughly
correspond to the NW and SE arms. Because the velocity distribution does not
suggest any rotating motion, the narrow components could represent a bipolar
outflow rather than a disk or a spherically expanding shell. The NW and SE arms
are blue- and red-shifted components, respectively, while the NW arm is fainter
than the SE one. It can not be explained with the inclination effect of the
nebula. Broad components in $\element[][12]{CO}$ and $\element[][13]{CO}$ only
appear around the central star with $\sim5\arcsec$ radii. It could reflect the
inner core, as seen in our NIR image. We will discuss more details of these
narrow and broad components in sect.\,\ref{nebula}.

\begin{table}
  \begin{center}
  \caption[]{Model parameters}
  \label{modelpar}
  \begin{tabular}{lcl} \hline
   parameters        & adopted values        & comments            \\
   \hline
   \multicolumn{3}{c}{stellar model}                               \\
   \hline
   $T_\mathrm{eff}$  & 2\,800~K              & by SED fit          \\
   $L_\star$         & 7\,000~$L_{\sun}$     & by SED fit          \\
   $d$               & 2.5~kpc               & adopted             \\
   $R_\star$         & 2.5$\times10^{13}$~cm & calculated          \\
   $\alpha_\star$    & 0.66~mas              & calculated          \\
   \hline
   \multicolumn{3}{c}{model geometry}                              \\
   \hline
   $R_\mathrm{in}$   & 7~$R_\star$           & by SED fit          \\
   $R_\mathrm{tr}$   & 50~$R_\star$          & by SED fit          \\
   $R_\mathrm{out}$  & 23\,000~$R_\star$     & adopted             \\
   $\tau$            & 20 ($V$), 1.5 ($H$)   & by SED fit          \\
   $\beta_1$         & 2                     & adopted             \\
   $\beta_2$         & 0.8                   & by SED fit          \\
   \hline
   \multicolumn{3}{c}{grain model}                                 \\
   \hline
   $a_\mathrm{min}$  & 0.005~$\mu$m          & adopted             \\
   $a_\mathrm{max}$  & 0.3~$\mu$m            & by polarization fit \\
   $n(a)$            & $a^{-3.5}$            & adopted             \\
   \hline
  \end{tabular}
  \end{center}
\end{table}

\section{Radiative transfer modeling}\label{modeling}
\subsection{Model assumptions}
We performed radiative transfer calculations to estimate the stellar parameters
and to investigate the dust properties in the envelope. We used a Monte Carlo
code, which simulates thermal emission and light scattering by dust grains in
a model geometry \citep{ohnaka06}. This code computes the SEDs and a dust
temperature distribution, and generates Stokes $IQUV$ images. Our code was
applied for interpretation of the mid-infrared spectro-interferometric data of
the silicate carbon star IRAS~08002--3803 \citep{ohnaka06} and the near-infrared
polarimetric data of the bipolar proto-planetary nebula (PPN) Frosty Leo
\citep{murakawa07}.

Although attempts to estimate the stellar parameters have been made, they are
still uncertain. \cite{nakashima04} have estimated a distance ranging between
2.5 and 3.9~kpc by applying the bolometric luminosity of 8\,000~$L_{\sun}$,
which is typical for AGB stars. The lower limit is obtained if the estimated
interstellar extinction of $A_V=16.3$~mag from their CO integrated intensity
($V_{lsr}=27$--33~km~s$^{-1}$ component) is assumed to be from a foreground
cloud. If it is from a background cloud, the upper limit is obtained. Following
\cite{nakashima04}, we apply the distance $d$ of 2.5~kpc in our modeling and
discuss the effect of the distance on the luminosity in
sect.\,\ref{stellar_prop} The stellar effective temperature is also not well
known, but is expected to be less than 4\,000~K, taking into account the
possible spectral type of a late K or an M giant (Wood, P. R.~2002, private
communication). The bolometric luminosity and the stellar temperature are
estimated by SED fitting.

The detection of several oxygen-rich molecular lines \citep[e.g. masers of SiO
and H$_2$O,][]{nd00} and a weak absorption feature at 10~$\mu$m \citep{volk91}
indicate an oxygen-rich environment in the envelope. Therefore, we assume bare
silicate grains. We use the optical constants of astronomical silicate
\citep{draine85}. To simplify the grain model, we applied a spherical shape
with an MRN-like power law size distribution; i.e., $a_\mathrm{min}\le a \le
a_\mathrm{max}$ and $n\left(a\right)=a^{-3.5}$, where $a$ and $n\left(a\right)$
are the grain radius and the size distribution function, respectively
\citep{mrn77}. In our Monte Carlo experiments, the size distribution averaged
values of the opacities and the scattering matrix components are used, which
are given by
\begin{eqnarray}
<X>=\int^{a_\mathrm{max}}_{a_\mathrm{min}}X\left(a\right)n\left(a\right)da\Big/\int^{a_\mathrm{max}}_{a_\mathrm{min}}n\left(a\right)da,\nonumber
\end{eqnarray}
where $X$ is the extinction cross section $C_\mathrm{ext}$, the absorption
cross section $C_\mathrm{abs}$, or the scattering amplitude matrix elements
$S_1\left(\theta\right)$ or $S_2\left(\theta\right)$. The averaged scattering
cross section $C_\mathrm{sca}$ is given by
$<$$C_\mathrm{sca}$$>=<$$C_\mathrm{ext}$$>$$-$$<$$C_\mathrm{abs}$$>$.
We set the minimum grain size $a_\mathrm{min}$ to be 0.005~$\mu$m since there
is no crucial clue to determine it. The maximum grain size $a_\mathrm{max}$ is
constrained by our modeling.

The \object{IRAS~19312+1950} nebula already shows a distinct and complex
bipolar appearance in the $\sim10\arcsec$ scale FOV. However, in our modeling,
we focus on estimating fundamental properties such as the stellar temperature,
the bolometric luminosity, the dust envelope mass, and the present-day
mass-loss rate instead of reproducing the ring-like structure. Because the inner
core ($r<3\arcsec$) has a nearly spherical structure, although an indication of
asymmetry is expected, we apply one-dimensional spherically symmetric
geometries, and such an approximation would satisfy our science goal.
The outer radius $R_\mathrm{out}$ is set to be a fixed value of
23\,000~$R_\star$, which corresponds to the extension of the nebula of
$\sim$15$\arcsec$ detected in the 2MASS $JHK$-band images, assuming the distance
of 2.5~kpc. Since two origins of dust (i.e., the stellar system and ambient
clouds) have been suspected before, we assume a model geometry with a radial
density profile given by $\rho\left(r\right)=
\rho_1\left(r/R_\mathrm{in}\right)^{-\beta_1}+\rho_2\left(r/R_\mathrm{in}\right)^{-\beta_2}$,
where $R_\mathrm{in}$ is the inner radius of the dust shell and the densities
of two components become equal at a transition radius $R_\mathrm{tr}$.
The values of $\rho_1$ and $\rho_2$ are determined from an input value of the
radial optical depth of the shell ($\tau$).

\begin{figure}
  \centering
  \includegraphics[width=8.5cm,keepaspectratio]{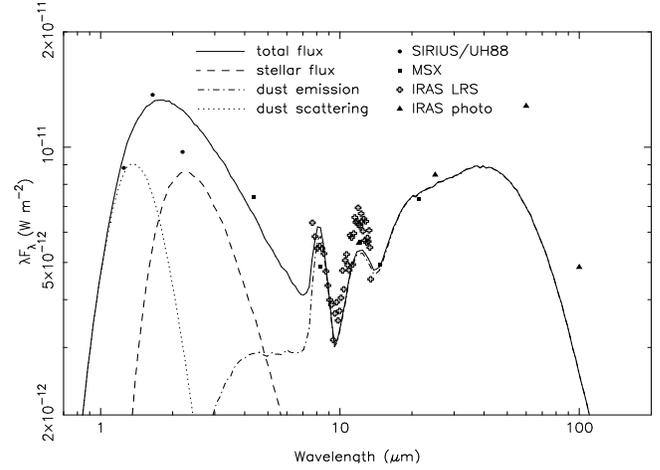}
  \caption{A comparison of the model SED with the observational results.
           The solid, dotted, dashed, and dashed-dotted curves denote the model
           results of the total flux, scattered light by dust in the envelope,
           the attenuated stellar flux, and the thermal emission from the
           envelope, respectively. The filled circles, filled squares, open
           crosses, and filled triangles are from the photometric results using
           SIRIUS on the UH88 telescope and MSX, the IRAS low resolution
           spectrum (LRS), and the IRAS photometry, respectively. The estimated
           interstellar extinction of $A_V=16.3$~mag is corrected in the SIRIUS
           data. The estimated background contributions of 30\,\% and 62\,\% of
           the 60 and 100~$\mu$m flux are subtracted from the IRAS photometry.
         }
  \label{sed}
\end{figure}

\begin{figure}
  \centering
  \includegraphics[width=8.5cm,keepaspectratio]{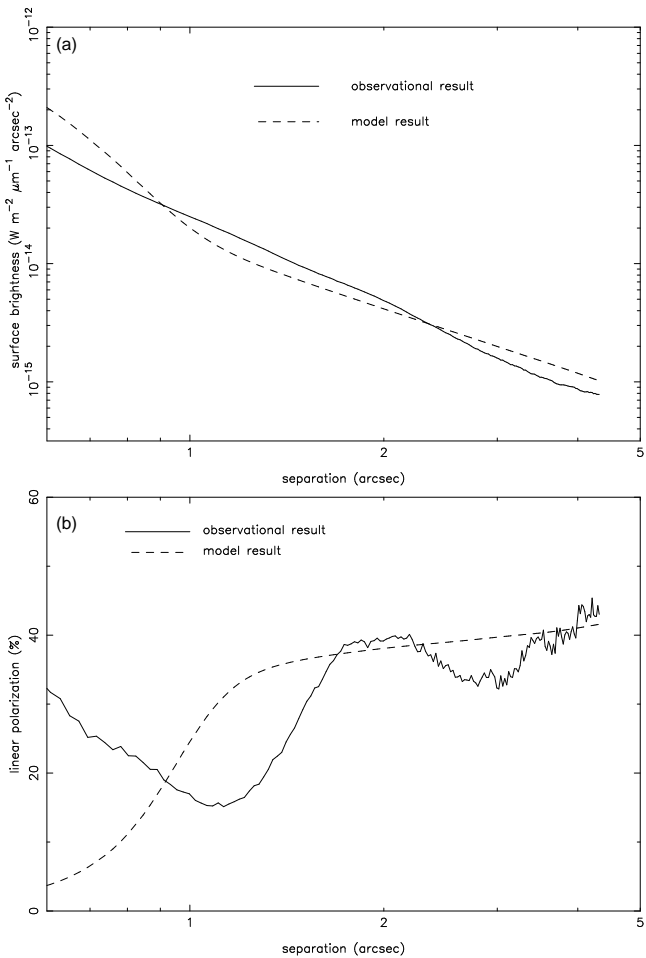}
  \caption{The radial profiles of the total intensity (a) and linear
           polarization (b), respectively. The solid curve and the dashed curve
           denote the observational result and the model result, respectively.
          }
  \label{profile}
\end{figure}

\subsection{Model results}\label{model_results}
We found that exponent values of $\beta_1=2$ and $\beta_2=0.8$ produce the best
fit of the mid-infrared SED, in particular the self-absorption silicate feature
around 10~$\mu$m. With respect to the grain properties, we obtained the maximum
grain size of $a_\mathrm{max}=0.3~(\pm0.05)~\mu$m, which satisfies the fraction
of linear polarization of our observational result. The extinction cross section
$C_\mathrm{ext}$, the dust albedo $\omega$, and the scattering asymmetry
$g$-parameter at 1.65~$\mu$m are $8.47\times10^{-7}$~cm$^2$, 0.389, and 0.187,
respectively. We searched the following parameter ranges: $T_\mathrm{eff}$ of
2\,500 to 3\,000~K with a 100~K step, $L_\star$ of 5\,000 to $10\,000~L_{\sun}$
with a $1\,000~L_{\sun}$ step, $R_\mathrm{in}$ of 3, 5, 7, and 10~$R_\star$,
$R_\mathrm{tr}$ of 10, 30, 50, and 100~$R_\star$, and $\tau_V$ of 7, 10, 20,
and 50. These combinations yield 2\,304 models, and we have examined their SEDs.
The best fit model has $T_\mathrm{eff}=2\,800~(\pm100)$~K,
$L_\star=7\,000~(\pm1\,000)~L_{\sun}$, $\tau_V=20$ (1.5 at 1.65~$\mu$m),
$R_\mathrm{in}=7~R_\star$, and $R_\mathrm{tr}=50~R_\star$. The stellar radius
$R_\star$ and its apparent size at 2.5~kpc are $2.5\times10^{13}$~cm and
0.66~mas, respectively. The model parameters are summarized in
Table~\ref{modelpar}.

The SED of the best model was compared with the observed SEDs in Fig.~\ref{sed}.
The data are from from $JHK$-band photometry obtained using the SIRIUS camera
on the University of Hawai'i 88-inch telescope (UH88), which is dereddened with
$A_V=16.3$~mag to correct for the foreground extinction \citep{nakashima04},
the Mid-course Space eXplorer (MSX) point source catalogue, the InfraRed
Astronomical Satellite (IRAS) low-resolution spectrum (LRS), and the IRAS
photometry. Because IRAS~19312+1950 is located near the Galactic plane, a large
fraction of flux in FIR is expected from background emission \citep{izumiura99}.
We obtain a rough estimation of the background contributions of 30\,\% at
60~$\mu$m and 62~\% at 100~$\mu$m from the IRAS Sky Survey Atlas (ISSA). Since
the pixel scale of the images is $1\farcm5$, the target is unresolved and seems
to be located in a region with complex emission features; the real values could
exceed the above values. These background contributions are subtracted from the
results of 60 and 100~$\mu$m IRAS photometry, plotted in Fig.~\ref{sed}.
Scattered light from the dust shell and the stellar flux dominates at
wavelengths shorter than 5~$\mu$m, while the thermal emission from the dust
shell does at longer wavelengths. The dust temperatures at the transition radius
(50~$R_\star$) and at the inner boundary of the shell (7~$R_\star$) are 290~K
and 960~K, respectively. The inner boundary temperature suggests a steady
formation of fresh dust in the outflow as oxygen-rich dust starts to condense
at temperatures typically in the range 1\,000--1\,500~K. The flux from the
inner shell mainly contributes at wavelengths shorter than 10~$\mu$m and the
flux from the outer shell at longer wavelengths. Our model SED is fair
agreement with the IRAS LRS 10~$\mu$m spectrum, which shows the silicate
self-absorption \citep{volk91}. The FIR flux of our model results are
sufficiently lower than the background subtracted results of the IRAS
photometry at 60 and 100~$\mu$m. The possible reasons are that our model
geometry (i.e., one-dimensional spherical model) is too simple and the
estimation of background contribution is too low, as already described.

The applied distance of 2.5~kpc in the above analysis is based on an assumption
that all CO emissions of $V_{lsr}=27$ -- 33\,km\,s$^{-1}$ component are from
foreground clouds \citep{nakashima04}. It is hard to determine the distance to
the object in general. However, it may be possible to be constrained by taking
the contribution of the interstellar extinction into account. If all CO
emissions are assumed to be from background, the distance to the object is
estimated to be 3.9~kpc \citep{nakashima04}. This results in all foreground
extinctions are due to the interstellar extinction of $A_V=6.2$ mag applying
the standard interstellar extinction \citep[e.g.][]{allen73,deguchi98} and the
optical depth of the envelope of $\tau_V=31$. We performed some modeling
with different contributions of the foreground extinction between 6.2 and 16.3
mag. We found that a model with $\tau_V>23$ (i.e., a foreground extinction
$A_V<14.9$ mag) fits the observed SED, and a large $A_V$ is preferred more than
a small $A_V$. Thus, it is likely that the object is near the back side edge or
behind ambient clouds. Then we take into account the contribution of the
interstellar extinction with a fixed foreground cloud component. We found good
solutions in SED fit when the interstellar extinction $A_V$ is between 2.2 and
6.7 mag. From this result, we estimate a possible distance to be between 1.4
and 4.2~kpc.

We also made Stokes $IQU$ images with another Monte Carlo experiment. Since the
thermal emission is negligible in the $H$-band, as seen in the SED
(Fig.~\ref{sed}), we simulate with monochromatic light at a wavelength of
$1.65~\mu$m to obtain Stokes images at a particular wavelength much more
efficiently than the full radiation transfer mode, which is applied to calculate
SEDs. In this fixed wavelength mode, a photon packet is always scattered at an
interacting point, and the weight of the photon packet is reduced by a factor
equal to the dust albedo $\omega$. The output Stokes images were convolved with
a Gaussian function with a 0\farcs6 diameter, which corresponds to the observed
PSF size in the $H$-band. Figure \ref{profile} (a) shows a comparison of radial
intensity (Stokes $I$) distribution of our model to the azimuthally averaged
observational result. Our observational result can be actually fit with
$\beta=1.2$ if a single power law is applied. However, it provides a poorer fit
of the 10~$\mu$m silicate self-absorption feature and too low flux in the FIR
and we could not find any adequate solutions to fit the observed SED and the
radial intensity distribution simultaneously with a single power law model.
Although we see somewhat steep and shallower slopes in our two component model
within $0\farcs9$ from the central star and further, respectively, the two
component model can reproduce the observed data reasonably well. Figure
\ref{profile} (b) shows the radial profile of the fraction of linear
polarization. Since there are regions with small polarizations along the
\textsf{S}-shaped arm, we omitted the data towards the central star within
$r<$~$1\farcs3$ because of contamination by unpolarized light from the central
star PSF. On the other hand, the linear polarization gently increases for
$r>1\farcs3$. Although our observational result of the radial profile has
a somewhat complex shape because of the \textsf{S}-shaped arm in the bipolar
nebula, the overall shape of our model result is in good agreement with our
observational data. Changing the maximum grain size does not affect the shape
of the radial profile much, but it does change the absolute fraction of linear
polarization dramatically. For a smaller $a_\mathrm{max}<0.25~\mu$m, a larger
polarization $>45~\%$ is attained, and for a larger $a_\mathrm{max}>0.37~\mu$m,
a lower polarization $<35~\%$ at $r=4\farcs5$ is attained.

\section{Discussion}\label{discussion}
The puzzling thing about IRAS~19312+1950 is that this object shows
characteristics of several object classes. The reported massive envelope
\citep{deguchi04} and an extremely high mass-loss rate are reminiscent of
a supergiant. However, some C- and N-bearing molecules such as NH$_3$, SO,
H$_2$CS, and CH$_3$OH \citep{deguchi04} in the envelope are ones often detected
in dark clouds. The detection of SiO (and H$_2$O) masers and a stronger OH maser
satellite line at 1612~MHz than main lines at 1665 and 1667~MHz suggest an AGB
star or a post-AGB star. Furthermore, an FIR emission distribution
($\ge1\arcmin$) extends sufficiently more than the nebulosity seen in the NIR
($\sim30\arcsec$). As already suspected, these characteristics are probably
caused by contributions from the matter originated outside the object; i.e.,
ambient clouds \citep{deguchi04,nd05}. Thus, we will discuss the properties of
the central star and the dust in the envelope separately.

\subsection{Stellar properties}\label{stellar_prop}
From our radiative transfer modeling, the stellar temperature and the distance
are estimated to be $T_\mathrm{eff}=2\,800$~K and $d=1.4$ -- 4.2~kpc,
respectively. The uncertainty of the distance causes the estimated luminosity
(1\,100$d^2~L_{\sun}$) to range between 2\,200\,$L_{\sun}$ and
19\,000\,$L_{\sun}$. With these properties, the object class of IRAS~19312+1950
could be classified in the Hertzsprung-Russell diagram. Considering YSOs, low
mass YSOs are much fainter \citep[$L_\star<10~L_{\sun}$, e.g.][]
{hartigan95,whitney04}, and luminous YSOs such as Herbig Ae$/$Be stars are much
hotter ($T_\mathrm{eff}>4\,000$~K) \citep[see also Fig.~1 of theoretical
pre-main sequence tracks in the HR diagram presented by][]{palla99}. Thus,
IRAS~19312+1950 is unlikely to be a YSO. Considering evolved stars, it is also
unlikely to be a supergiant because the estimated luminosity of IRAS~19312+1950
is within the AGB limit \citep[$\sim5\times10^4~L_{\sun}$,][]
{paczynski71,wood83} and is sufficiently lower than several known supergiants;
e.g. $7\times10^5~L_{\sun}$ for \object{IRC +10 420} \citep{jones93},
$2\times10^5~L_{\sun}$ for \object{NML Cyg} \citep{bloecker01},
$5.5\times10^5~L_{\sun}$ for \object{VY CMa} \citep{sopka85}. Therefore,
the possibility remains that IRAS~19312+1950 is an AGB or a post-AGB star.

From our modeling, we estimate the dust envelope mass to be $0.14~M_{\sun}$ if
the distance to the object of 2.5~kpc is applied. Comparing this with a gas
envelope mass of $\sim$25~$M_{\sun}$ \citep{deguchi04}, the gas-to-dust mass
ratio is found to be $\sim$$180:1$, which is very typical for the neutral
interstellar medium in the galactic plane and the winds of AGB stars of about
solar metallicity \citep{knapp74,knapp85}. We also estimate a mass-loss rate.
While the value of $2.6\times10^{-4}~M_{\sun}$~yr$^{-1}$ obtained by
\cite{deguchi04} is derived at $r\sim3\arcsec$ from the central star,
our modeling allows us to derive the current value. Applying the above
gas-to-dust mass ratio and the expansion velocity of 25~km~s$^{-1}$, which is
approximately half the width of the CO $J=1$--0 emission velocity distribution
\citep{nakashima04}, we obtain the current mass-loss rate of
$5.3\times10^{-6}~M_{\sun}$yr$^{-1}$. This value is in good agreement with
the value of $6.3\times10^{-6}~M_{\sun}$yr$^{-1}$ calculated with an empirical
formula of mass-loss rate
log$\dot{M}=-5.65+1.05$~log$\left(L/10\,000~L_{\sun}\right)
-6.3$~log$\left(T_\mathrm{eff}/3\,500~\mathrm{K}\right)$
for M-type AGB stars \citep{vanloon05}. From the above properties, the central
star of IRAS~19312+1950 shows characteristics of dust-enshrouded AGB stars.

\subsection{Nebula and duality of the dust chemistry}\label{nebula}
Dust shells formed in the AGB phase often have a nearly spherical shape, as
seen in \object{TT\,Cyg} \citep{olofsson00} and in large field of view
($>10\arcsec$) images of \object{IRC\,+10\,216} \citep{mh00}. Bipolarity,
which is seen in a large fraction of PNs \citep{corradi95}, is thought to be
formed by the fast-wind during the PPN phases \citep[e.g. review by][]{bf02}.
However, the evidence for the presence of high velocity jets and a rotating disk
have been reported in some AGB stars, which are not thought to undergo the
intensive mass loss that forms an aspherical dust shell; e.g., \object{V Hya}
\citep{kahane96,sahai03,hirano04}, \object{X Her} \citep{kj96,nakashima05},
and \object{RV Boo} \citep{bergman00}. These facts are hard to explain with
a scenario of stellar evolution of a single star, and it is thought that binary
companion(s) must play an important role in shaping aspherical shells
\citep[e.g.][]{soker97,soker02}. With respect to IRAS~19312+1950, the
\textsf{S}-shaped arm is detected in NIR images, and such a morphology might
be a result of interactions with binary companions. However, while the high
velocity broad components are extended and seen as a bipolar outflow, and
narrow components concentrate close to the central star in case of V Hya,
the kinematic properties in IRAS~19312+1950 is opposed sense. A possible
interpretation is given as follows: the object is probably embedded in an
ambient cloud. The fast wind from the central star, which is detected in the
broad components in \element[][12]{CO} and \element[][13]{CO}, rapidly slows
down due to the thick ambient cloud matter. As a result, the narrow components
of 35--36 and 37--38~km~s$^{-1}$ in \element[][13]{CO} $J=1$--0 extending
$\sim15\arcsec$ towards the NW and SE directions are seen.

Duality of dust chemistry, i.e., detection of C- and O-bearing molecules, in
the IRAS~19312+1950 nebula is also an interesting issue. Carbon stars with
oxygen-rich molecules in the envelope, such as silicate carbon stars
\citep{little-marenin86,willems86} and the \object{Red Rectangle}
\citep{waters98}, are known or are expected to have binary companions.
The most widely accepted hypothesis of the presence of oxygen-rich dust in
envelopes around these carbon stars suggests that silicate was ejected by mass
loss from the primary star and was trapped in a circumbinary disk or a companion
disk when the primary carbon star was an M giant \citep{morris87,lloyd-evans90}.
Evidence for the presence of a disk or a reservoir is obtained with detection
of a narrow velocity component ($v<5$~km~s$^{-1}$) in CO emission lines
\citep[e.g.][]{kahane98}. With respect to IRAS~19312+1950, the detection of
SiO maser emissions \citep{nd00} presents an important constraint on the nature
of the central star. SiO maser emissions have been detected toward evolved
stars with only a few exceptional star forming regions; e.g., \object{Orion-KL}
\citep{baud80} and \object{W51 IRS2} \citep{fuente89} and are produced above
the surface (5--10~AU or a few $R_\star$) of evolved stars \citep{elitzur80}.
To date, no detection has been reported in a silicate carbon star
\citep[e.g.][]{nakada87} or in any other carbon star \citep[e.g.][]
{lepine78,schoier06}. Thus, IRAS~19312+1950 is probably an oxygen-rich star at
least at this moment and C-bearing molecules do not originate in the mass loss
from the central star. The origin of C- and N-bearing molecules, which are
detected towards IRAS~19312+1950, could be chemical reactions in the wind
\citep{lindqvist88,wm97,duari99} or ambient cloud matter, which kinetically
merges with the matter ejected by the mass loss from the central star.

\section{Conclusion}
We present $H$-band imaging polarimetry of the peculiar bipolar nebula
IRAS~19312+1950 using CIAO on the Subaru telescope. The total intensity image
clearly shows a point-symmetric \textsf{S}-shaped arm extending towards the
northwest and the southeast. The polarization map revealed a centro-symmetric
polarization vector pattern in the entire bipolar lobes with polarizations of
30--60~\%. These results indicate that the nebulae are associated with the
central star and are seen in light scattered by dust in the nebula. We find low
polarizations ($P<20~\%$) on the east side of the SE and NW arms, suggesting
that the \textsf{S}-shaped arm has a physically ring-like structure instead of
a hollow or cavity structure in the lobes.

We also investigated physical properties of the central star and the nebula by
means of dust radiative transfer calculations. The estimated stellar
temperature and bolometric luminosity are 2\,800~K and 7\,000~$L_{\sun}$,
respectively. The distance uncertainty is found to be between 1.4 and 4.2~kpc.
Comparing the gas envelope mass of 25~$M_{\sun}$ to the dust envelope mass of
0.14~$M_{\sun}$, the gas-to-dust mass ratio obtained is $180:1$. With this
value, the current mass-loss rate is estimated to be
$5.3\times10^{-6}~M_{\sun}$yr$^{-1}$. This is in good agreement with a value of
$6.3\times10^{-6}~M_{\sun}$yr$^{-1}$ calculated with an empirical mass-loss
formula for M-type AGB stars. From above stellar properties and detections
of SiO, H$_2$O, and OH masers, the central star of IRAS~19312+1950 shows
characteristics of an oxygen-rich, dust-enshrouded AGB star rather than a YSO
or a supergiant.

Our results support the idea proposed by \citep{deguchi04,nd05} that the object
is embedded in ambient clouds, because the large envelope mass of 25~$M_{\sun}$
is impossible to explain with only mass loss from the possible AGB central star.
Although the detected C- and N-bearing molecules can form by chemical reactions
in the oxygen-rich envelope, it could also originate from ambient cloud matter,
which kinetically merges with the outflow of the central star.

\begin{acknowledgements}
We would like to thank referee Dr. van Loon, J. Th. for his particularly
thoughtful comments that led to a significant improvement of this paper.
We would also like to thank Drs. B. M. Lewis and P. R. Wood for providing us
with observational results of OH masers and $K$-band spectra of IRAS~19312+1950,
respectively.

\end{acknowledgements} 



\end{document}